\documentclass[prl,twocolumn,showpacs,amsmath,amssymb,superscriptaddress]{revtex4-1}

\usepackage{graphicx}
\usepackage{amssymb,amsmath}
\usepackage{dcolumn}
\usepackage{bm}
\usepackage{mathrsfs}
\usepackage{color}
\usepackage{hyperref}

\begin{document}

\title{Landau versus Spin Superfluidity in Spinor Bose-Einstein Condensates}

\author{H. Flayac}
\affiliation{Institute of Theoretical Physics, \'{E}cole Polytechnique F\'{e}d\'{e}rale de Lausanne EPFL, CH-1015 Lausanne, Switzerland}
\author{H. Ter\c{c}as}
\affiliation{Institut Pascal, PHOTON-N2, Clermont Universit\'{e}, Blaise Pascal University, CNRS, 24 avenue des Landais, 63177 Aubi\`{e}re Cedex, France}
\author{D. D. Solnyshkov}
\affiliation{Institut Pascal, PHOTON-N2, Clermont Universit\'{e}, Blaise Pascal University, CNRS, 24 avenue des Landais, 63177 Aubi\`{e}re Cedex, France}
\author{G. Malpuech}
\affiliation{Institut Pascal, PHOTON-N2, Clermont Universit\'{e}, Blaise Pascal University, CNRS, 24 avenue des Landais, 63177 Aubi\`{e}re Cedex, France}

\begin{abstract}
We consider a spin-$1/2$ Bose-Einstein condensate prepared initially in a single spin projection. The two channels of excitations existing in such a system (namely density and spin waves) are discussed and we show how pure spin waves can be excited in the presence of local magnetic defects. We analyze the role played by spin excitations on the Landau superfluidity criterion and demonstrate the absence of absolute superfluidity for the antiferromagnetic condensate.  In the ferromagnetic case, we identify two critical velocities for the breakdown of superfluidity.
\end{abstract}
\maketitle

\emph{Introduction.---} Superfluidity \cite{50years} is undoubtedly the most famous and striking phenomenon linked with Bose-Einstein condensation (BEC), since it corresponds to regimes in which the condensate is stable against the creation of excitations. According to the Landau criterion, it occurs whenever the condensate propagates at a velocity smaller than the speed of its excitations \cite{PitaevskiiBook}. The latter are sound (density) waves, or bogolons \cite{Bogoliubov}, characterized by a linear dispersion for long wavelengths and a density-dependent celerity of sound. When the internal spin degree of freedom of the particles is taken into account, the situation becomes even more interesting. Indeed, spinor BECs \cite{ReviewUeda1,OpticalTrap}, which are not restricted to atomic systems but also include other systems such as exciton-polaritons \cite{BECPolaritons,ReviewSpin,Tercas} and magnons \cite{BECMagnons,MagnonSpinor}, have already brought remarkable discoveries, such as the formation of topological defects, exotic phases and spin textures \cite{ReviewUeda2,SpinDomains}. The physics of spinor BECs becomes even richer in the presence of spin-orbit interactions \cite{SOBECExp,SOBEC,ExoticSF,ShelykhPRL2006,NOSHE}.

In general, superfluidity criteria for spinor BECs are less trivial, as the critical velocity is not uniquely defined \cite{ReviewUeda1}. As a consequence, spin and density degrees of freedom are mixed and the channel associated with pure spin excitations \cite{SpinWaves1,SpinWaves2,SpinWaves3}, in analogy to magnetism \cite{SWMagnetic}, is harder to identify whereas the task of isolating, protecting, and controlling pure spin excitations would be of great interest for various applications. The study of such systems leads to the understanding of spin superfluidity as a phenomenon distinct from the usual Landau superfluidity: the onset of a spin-polarized (or magnetized) flow absent of spin excitation. In fact, spinor BECs possess undeniable similarities with magnetic systems: depending on the relative strength between the intra and inter-component interactions (tunable e.g. via Feshbach resonance \cite{Feshbach} in the atomic case), the system can exhibit either ferromagnetic \cite{Ferromagnetic}, antiferromagnetic \cite{Antiferromagnetic}, paramagnetic \cite{Paramagnetic} or even diamagnetic features \cite{SpinMeissner}.

In this Letter, we discuss the spin-density separation in magnetized spin-$1/2$ BECs  
and investigate its superfluid features. As it is known, spin-density separation, equivalent to spin-charge separation in Coulomb chains for neutral systems \cite{Giamarchi}), can occur in BEC at arbitrary dimensions \cite{Chung}. Here, we consider the separation in spinor condensates where the magnetization is either found in the ground state (ferromagnetic condensate) or brought by an applied magnetic field (antiferromagnetic condensate). By showing that the spin and density excitations can decouple, we argue that a spin current, protected against spin excitations, is possible for a wide range of parameters. As a remarkable feature of the spin-density separation, we show that the complete suppression of spin excitations is possible even in the supersonic regime, where sound waves can develop. We test the spin superfluidity criterion by simulating the formation and suppression of spin waves in a condensate past a magnetic defect and establish the corresponding magnetic drag force.

\emph{The spin-$1/2$ condensate.---}
We shall consider a spinor BEC having two allowed spin projections on the $z$ quantization axis. At the mean field level, the homogeneous system is governed by the energy functional
\begin{equation}\label{Ef}
E = \int {d\mathbf{r} } \left[ \begin{array}{r}
\sum\limits_{j  =  1,2 } {\psi _j ^*\left( { - \frac{{{\hbar ^2}{\Delta}}}{{2m}} + \frac{{{g_{jj}}}}{2}{{\left| {{\psi _j }} \right|}^2}} \right){\psi _j }} \\
 + {g_{12}}{\left| {{\psi _ 1 }} \right|^2}{\left| {{\psi _ 2 }} \right|^2}-\mathbf{H} \cdot \mathbf{S}
\end{array}\right],
\end{equation}
written for the spinor
\begin{eqnarray}
{\boldsymbol{\Psi }}\left( {\mathbf{r},t} \right) = \left( \begin{array}{l}
{\psi _ 1 }\left( {\mathbf{r},t} \right)\\
{\psi _ 2 }\left( {\mathbf{r},t} \right)
\end{array} \right) = \left( \begin{array}{l}
\sqrt {{n_ 1 }\left( {\mathbf{r},t} \right)} {e^{i{\theta _ 1 }\left( {\mathbf{r},t} \right)}}\\
\sqrt {{n_ 2 }\left( {\mathbf{r},t} \right)} {e^{i{\theta _ 2 }\left( {\mathbf{r},t} \right)}}
\end{array} \right)
\end{eqnarray}
in the Madelung representation, where $n_{j}$ and $\theta_{j}$ respectively represent the density and phase of the $j=1,2$ spin projection. Here, $\boldsymbol{H}=(H_x, H_y, H_z)^T$ is a generic (effective) magnetic field. Its $H_z$ component corresponds to the action of a magnetic field inducing a Zeeman splitting while the transverse components $H_{x,y}$ mixing the spin states could be produced e.g. by a microwave field for atoms \cite{PethickBook} or a polarization splitting for polaritons \cite{Klopo}. $g_{jj}$ and $g_{12}$ are the strength of the intra- and inter-spin interactions, respectively and in the following, we shall consider $g_{11}=g_{22}=g$. The pseudospin vector $\mathbf{S}$ allows the mapping of our problem to a magnetic system \cite{SpinTextures} and its components are linked to the spinor $\boldsymbol{\Psi}$ via the identities
\begin{eqnarray}
\label{SxId}
{S_x} &=& \frac{1}{2}\left( {{\psi _ 1 }\psi _ 2 ^* + \psi _ 1 ^*{\psi _ 2 }} \right) = \sqrt {{n_ 1 }{n_ 2 }} \cos \left( {\Delta \theta } \right)\nonumber \\
\label{SyId}
{S_y} &=& \frac{i}{2}\left( {{\psi _ 2 }\psi _ 1 ^* - \psi _ 2 ^*{\psi _ 1 }} \right) = \sqrt {{n_ 1 }{n_ 2 }} \sin \left( {\Delta \theta } \right)\\
\label{SzId}
{S_z} &=& \frac{1}{2}\left( {{n_ 1 } - {n_ 2 }} \right), \nonumber
\end{eqnarray}
where $\Delta\theta=\theta_1-\theta_2$ is the relative phase. The $S_z$ projection can thus be seen as the magnetization of the system. The free energy of the condensate reads
\begin{eqnarray}\label{FreeEnergy}
\nonumber F =  &-& \mu {n} + \left( {g + g_{12}} \right)\frac{{n^2}}{4}\\
 &-& {H_x}{S_x}-{H_y}{S_y}-({H_z}-\Delta g S_z){S_z},
\end{eqnarray}
where $n=n_1+n_2$, $\Delta g=g-g_{12}$ and $\mu$ is the global chemical potential. We emphasize that only $\mu$ is conserved in Eq.(\ref{FreeEnergy}), differing from the case of a condensate mixture where the partial chemical potentials $\mu_{1,2}=g n_{1,2}+g_{12}n_{2,1}$ are simultaneously conserved. Moreover, we should avoid the energetic instability region $g_{jj}<0$ and consider only the case where the homogeneous magnetized ground state is stable. The last term contains an extra (interaction-induced) Zeeman splitting $H_\textrm{Ze}=-\Delta g S_z$ that can be seen as the spin-spin interaction. The homogeneous ground state is found from the minimization of $F$ solving $\partial_n F=0$, at fixed total density $n$, with respect to the pseudospin components given the normalization condition $S_x^2+S_y^2+S_z^2=n^2/4$. Let us focus on the case where $H_{x,y}=0$ treated as perturbations in the following.

In the case where $\Delta g>0$, the condensate is said to be \emph{antiferromagnetic} since its interaction energy is minimized by superimposing particles of opposite spin. The ground state has consequently no magnetization, namely $S_{z}=0$. The transverse components are undetermined according to $S_{x}^{2}+S_{y}^{2}=n^2/4$ and the chemical potential reads $\mu=(g+g_{12})n/2$. Provided that $\Delta g\neq0$, a metastable state is obtained if the condensate is prepared initially with some non-zero magnetization ($S_z\neq0$), thanks to the intrinsic Zeeman splitting $H_\textrm{Ze}$ that induces an effective magnetic field locking the magnetization.

By contrast, the \emph{ferromagnetic} condensate defined by $\Delta g<0$ has a naturally magnetized ground state given by $\mathbf{S}=(0,0,\pm n/2)^T$ and the chemical potential $\mu=g n$.

In the presence of an applied magnetic field $\mathbf{H}=H_z \mathbf{e}_z$, the antiferromagnetic condensate demonstrates some critical behavior due to the competition between $H_z$ and $H_\textrm{Ze}$. This effect is responsible for the so-called spin Meissner effect \cite{SpinMeissner}. Consequently only under the condition $H_z>H_{Ze}$ can the ground state become magnetized $\mathbf{S}=(0,0,n/2)^T$ and we assist to a transition towards a ferromagnetic condensate with $\mu=g n-H_z$.

\begin{figure}[ht]
\includegraphics[width=0.49\textwidth,clip]{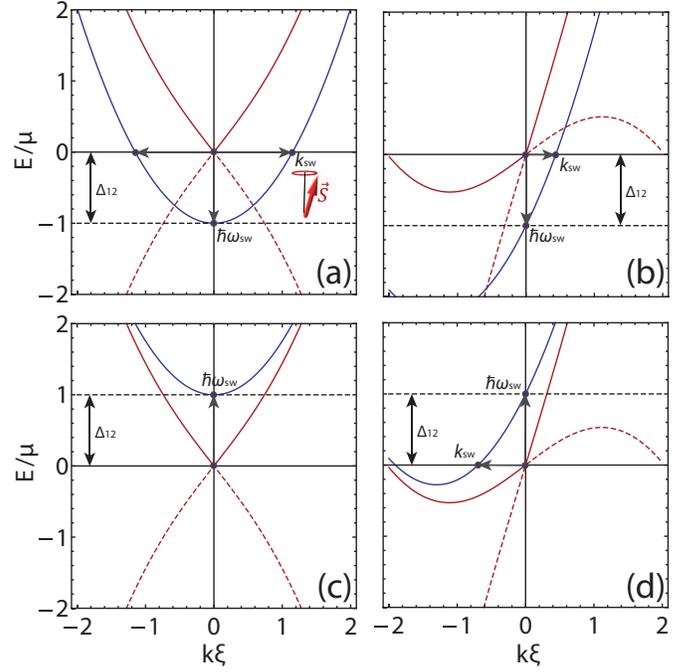}\\
\caption{(Color online) Dispersion of the elementary excitations of the magnetized condensate. The solid (dashed) red line shows the $E_1^{(+)}$ ($E_1^{(-)}$) Bogoliubov branches of the condensate and the solid blue line is the dispersion of the uncondensed component $E_2$ or the spin excitations channel. The horizontal (vertical) arrows point the excitation of spatial (temporal) spin waves. (a),(b) Antiferromagnetic and (c),(d) ferromagnetic condensate (a),(c) $v<c_m$ and (b),(d) $v>c_m$.}
\label{fig1}
\end{figure}

\emph{Spin-Density separation and pure spin waves.---}
Let us now describe the elementary excitations on top of the magnetized condensate $(S_x^0,S_y^0,S_z^0)^T=(0,0,+n_0/2)^T$. The minimization of the energy functional (\ref{Ef}) leads to the Gross-Pitaevskii equations:
\begin{eqnarray}
\label{GP1}
\nonumber i\hbar \frac{{\partial {\psi _ 1 }}}{{\partial t}} =  &-& \frac{{{\hbar ^2} }}{{2m}}{\Delta\psi _ 1 } -\mu\psi_1 + {g}{\left| {{\psi _ 1 }} \right|^2}{\psi _ 1 }\\
                                                        &+& {g_{12}}{\left| {{\psi _ 2 }} \right|^2}{\psi _ 1 }-\frac{{{H_z}}}{2}\psi_1,\\
\label{GP2}
\nonumber i\hbar \frac{{\partial {\psi _ 2 }}}{{\partial t}} =  &-& \frac{{{\hbar ^2} }}{{2m}}{\Delta\psi _ 2 } -\mu\psi_2 + {g}{\left| {{\psi _ 2 }} \right|^2}{\psi _ 2 }\\
                                                         &+& {g_{12}}{\left| {{\psi _ 1 }} \right|^2}{\psi _ 2 } +\frac{{{H_z}}}{2}\psi_2.
\end{eqnarray}
We start from the Bogoliubov ansatz \cite{Bogoliubov}
\begin{eqnarray}
\label{Ansatz1}
{\psi _ 1 } &=& {\sqrt {n_0}  + {A_ 1 }{e^{+i\left( {\mathbf{k}\cdot\mathbf{r} - \omega t} \right)}} + B_ 1 ^*{e^{ - i\left( {\mathbf{k}\cdot\mathbf{r} - {\omega ^*}t} \right)}}}\nonumber \\
\label{Ansatz2}
{\psi _ 2 } &=& {{A_ 2 }{e^{ i\left( {\mathbf{k}\cdot\mathbf{r} - \omega t} \right)}}},
\end{eqnarray}
that involves two counter-propagating plane waves in the condensate characterized by the weak amplitudes $A_1$ and $B_1$, the wavevector $\mathbf{k}$ and frequency $\omega$. We obviously seek for single plane waves of amplitude $A_2$ in the empty component. Injecting (\ref{Ansatz1}) into Eqs.(\ref{GP1},\ref{GP2}), the chemical potential $\mu={g}{n_0}-H_z$ is recovered and we obtain the following modes
\begin{eqnarray}
\label{Disp1}
E_1 ^{\left( \pm \right)}\left(k\right) &=&  \pm \sqrt {E_0\left(k\right) \left({E_0\left(k\right) + 2{\mu }} \right)}\\
\label{Disp2}
E_2 \left(k\right) &=&  E_0\left(k\right) + (H_z-\Delta g n_0),
\end{eqnarray}
where $E_0\left(k\right)=\hbar^2k^2/2m$. The elementary excitations on top of the condensed component are Bogolons or sound waves following the dispersion $\tilde{E}_1^{(\pm)}=\pm c_s |k|$ at low momentum, where $c_s=\sqrt{\mu/m}$ is the speed of sound. Eqs.(\ref{Disp1},\ref{Disp2}) are a signature of the separation between density and spin excitations. As a result of the spin-density separation, the static condensate is stable against the creation of these excitations, that have a positive $(|A_1|^2-|B_1|^2)E_1^{(+)}$ contribution to the energy, and is therefore superfluid in the Landau picture. The coefficients $A_1$ and $B_1$ can be explicitly found provided the normalization condition $|A_1|^2+|B_1|^2+|A_2|^2=1$, thus yielding
\begin{equation}
\label{AB}
{\left( {A_ 1 ^{\left( \pm \right)},B_ 1 ^{\left( \pm \right)}} \right)} = \frac{1}{{\sqrt {{\mu ^2} + \Delta {E^2}} }}{\left( {\mu , \mp \Delta E} \right)},
\end{equation}
where $\Delta E=E_0(k)+\mu-E _ 1 ^{(+)}$. The parabola of the second spin component defined by Eq.(\ref{Disp2}) is shifted by a quantity $\Delta_{12}=H_z-\Delta g n_0$ due both to the presence of the condensate and the Zeeman splitting. In Fig.\ref{fig1}(a,c) we depict illustrate the features of the dispersions (\ref{Disp1},\ref{Disp2}). Interestingly, the excitation of a plane wave $\delta\psi_2(x,t)=A_2\exp(i \mathbf{k} \cdot\mathbf{r} -\omega t)$, e.g. due to fluctuations or to the presence of a transverse field, will drive the pseudospin dynamics according to
\begin{eqnarray}
\label{SxSW}
S_x &=& \frac{{{A_ 2 } }}{2}\cos \left( {\mathbf{k}_{{sw}}\cdot\mathbf{r} - \omega_{{sw}} t} \right),\nonumber \\
\label{SySW}
S_y&=& \frac{{{A_ 2 } }}{2}\sin \left( {\mathbf{k}_{{sw}}\cdot\mathbf{r} - \omega_{{sw}} t} \right), \quad \mbox {and}\\
\label{SzSW}
S_z &=& \frac{{n_0^2 - \left|A_2\right| ^2}}{2},\nonumber
\end{eqnarray}
namely to a spatio-temporal modulation of the relative phase $\Delta\theta$ or the Larmor precession of $\mathbf{S}$ about the $z$-axis [see red arrow in Fig.\ref{fig1}(a)]. The magnetization is reduced by $\left|A_2\right|^2/2$ but does not fluctuate. Due to the spin-density separation, pure \emph{spin waves} can develop, in analogy to what happens in magnetic systems \cite{SWMagnetic}.

\begin{figure}[ht]
\includegraphics[width=0.5\textwidth,clip]{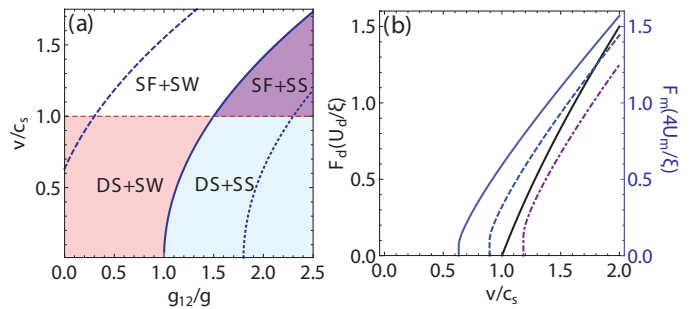}\\
\caption{(Color online) (a) Phase diagram of the propagating condensate. SF: Supersonic Flow, SW: Spin Waves formation, SS: Spin Superfluid and DS: Density Superfluid. The dashed, solid and dotted blue lines mark the frontier $v=c_m$ in the cases $H_z=\left\{-0.8,0,1.2\right\}mc_s^2$ respectively. The horizontal dashed line stands for $v=c_s$. (b) Drag forces experienced by the condensate. The blue, dashed blue and dashed-dotted purple lines depict the magnetic drag forces (right scale) $F_m$ for $H_z=\left\{0,1.4,0.5\right\}mc_s^2$ and $\alpha_2=\left\{1.2,0,1.2\right\}$ respectively. The solid black line show the mechanical drag force $F_m$ (left scale) for comparison. $\xi=\hbar/\sqrt{m\mu}$ is the healing length.}
\label{fig2}
\end{figure}

For both anti- and ferromagnetic condensates, pure temporal spin waves can always be excited by a transverse field that is local in time. They correspond to vertical transitions between the condensate and the parabola that become enhanced as the coupling is increased or the separation $\Delta_{12}$, defining the wave frequency $\hbar\omega_{{sw}}$, is reduced [see vertical arrows in Fig.\ref{fig1}]. However, as one can see from Fig.\ref{fig1}(a,c), the formation of spatial spin waves, requiring a perturbation that is local in space, is strongly dependent on the antiferromagnetic or ferromagnetic character of the condensate. Indeed, in the former case ($\Delta_{12}<0$), the parabola is redshifted [Fig.\ref{fig1}(a)] and therefore the horizontal transitions at the energy of the condensate are allowed, thus exciting a spin wave of wavevector ${k_{{sw}}}=\pm\sqrt{2m\Delta_{12}}/\hbar$ (horizontal arrows). In the latter case, the parabola is blueshifted [Fig.\ref{fig1}(c)] and there is no available state for the horizontal transitions namely spatial spin waves are fully suppressed.

Let us now consider that the condensate propagates at speed $\mathbf{v}$. In that case, the dispersion branches (\ref{Disp1},\ref{Disp2}) experience the Doppler shift $E_D(k)=\hbar \mathbf{v}\cdot\mathbf{k}$ [Figs.\ref{fig1}(b,d)]. According to the Landau criterion, superfluidity breaks down in the supersonic regime $|\mathbf{v}|>c_s$ where sound waves are excited. 

The vertical splitting $\Delta_{12}$ is independent of the condensate propagation speed, which simply means that there is no velocity dependence on the frequency or the amplitude of temporal spin waves. Nonetheless, the amplitude of the spatial spin waves is enhanced as the relative momentum is reduced according to $\hbar {k_{{{sw}}}}/m =  \pm \sqrt {2{\Delta _{12}}/m + v}  - v$, for an antiferromagnetic condensate [see Fig.\ref{fig1}(b)]. The situation is even more interesting for the ferromagnetic case, since spin waves can develop only above the critical speed
\begin{equation}\label{cm}
{c_m} = \sqrt {\frac{{2\left( {{H_z} - \Delta g{n_0}} \right)}}{m}}  = \sqrt {\frac{{{\mu _m}}}{m}}
\end{equation}
The natural extension of Landau's criterion therefore predicts absolute superfluidity (in both density and spin sectors) if the conditions $v<c_s$ (density superfluidity) and $v<c_m$ (spin superfluidity) are simultaneously fulfilled. Noteworthy is that in the general case $c_s\neq c_m$, allowing two hybrid superfluid behaviors in which one single type of excitation occurs namely if $c_s<v<c_m$ (density waves) or $c_m<v<c_s$ (spin waves). These features are summarized in the phase diagram depicted in Fig.\ref{fig2}(a). We can easily identify four distinct region [see captions], depending on which kind of excitations are allowed to develop given $\mathbf{v}$ and the ration $g_{12}/g$.

\emph{Spin wave generation.---}
The most current experiment to highlight superfluidity involves a condensate propagating against a potential barrier. As it is known, in the supersonic regime $v>c_s$, backscattered density waves appear upstream from the defect. In the subsonic regime, however, no acoustic radiation takes place. It is therefore natural to ask what happens to the spinor BEC flow past a "magnetic defect", which we consider to be a weak, local transverse field $\mathbf{H}_{||}(\mathbf{r})=H_x(\mathbf{r})\mathbf{e}_x$. The latter enters the dynamics via the coupling terms $-H_x(\mathbf{r})\psi_\mp/2$ in Eqs.(\ref{GP1},\ref{GP2}), respectively.

\begin{figure}[ht]
\includegraphics[width=0.5\textwidth,clip]{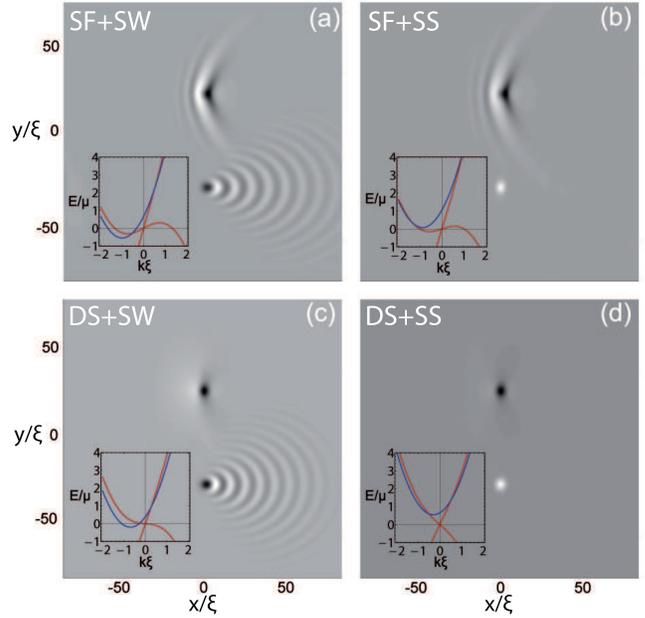}\\
\caption{(Color online) The four phases of the ferromagnetic condensate passing a barrier (upper) and a magnetic defect (lower) within the same flow. The colormap shows $|\psi_1 + \psi_2|^2$ and the insets display the dispersions. The visual order of the phases is preserved with respect to Fig.\ref{fig1}(b) for clarity.}
\label{fig3}
\end{figure}

We have performed numerical simulations considering a two-dimensional homogeneous, magnetized condensate propagating with the velocity $\mathbf{v}=v_x \mathbf{u}_x$, by setting the initial condition $\boldsymbol\Psi_0\left(\mathbf{r}\right)=(\sqrt {{n_0}} {e^{i{k}x}},0)^T$, with $k_x=m/\hbar v_x$. In order to distinguish between density- and spin superfluid situations, we have plugged both a potential barrier and a magnetic defect, spatially separated from each other, to respectively excite density and spin waves in the flow. The results are summarized in Fig.\ref{fig3}, where the four regions of the phase diagram of Fig.\ref{fig2}(a) are illustrated. We enclosed snapshots of the linear combination $|\psi_1+\psi_2|^2$, which allows to visualize density and spin waves simultaneously. In each panel, the insets show the corresponding dispersion of elementary excitations. Panel (a) corresponds to a supersonic flow configuration in which spin waves can develop as well ($v>c_s, c_m$). Interestingly, while the density waves form upstream from the barrier, the spin waves are found downstream due to their positive group velocity \cite{Note} as deduced Fig.\ref{fig1}(b),(d) and from the insets of Fig.\ref{fig3}. The panel (b) shows the exclusive spin superfluid regime in a supersonic flow, where only density excitations occur ($c_s<v<c_m$) and panel (c) illustrates the exclusive Landau superfluidity, for which only spin waves can be excited ($c_m<v<c_s$). Finally, panel (d) demonstrates the absolute superfluidity of the spinor condensate, in which no excitations are generated ($v<c_s,c_m$).

Finally, to further understand the differences between density and spin superfluidity, we address the question of dissipation in the spin flow. In the presence of a magnetic defect, the drag force is defined as $\mathbf{F}_m=-\int (\psi_1^*\bm \nabla H_x\psi_2+\mbox{c.c} )d\mathbf{r}$. For simplicity, we consider a localized defect of the type $H_x=U_m\delta(\mathbf{r})$, such that $\mathbf{F}_m=U_d\int (\psi_1\bm \nabla \psi_2 \vert_{\mathbf{r}=0}+\mbox{c.c})d\mathbf{r}$. Expanding the wave functions according to (\ref{Ansatz2}), and using the Hopfield amplitudes (\ref{AB}) modified by the Doppler shift $E_D(k)$, we obtain, in first order in $A_{1,2}$ and $B_1$, $\mathbf{F}_m=U_m n_0\int i\mathbf{k}S_f(k,\mathbf{k}\cdot \mathbf{v}) d\mathbf{k}/(2\pi)^2$, where
\begin{equation}
S_f(k,\omega)=\frac{1}{2m}\frac{k^2}{(\omega+i0)^2-\omega_-^2}
\label{SF}
\end{equation}
represents the dynamic structure factor associated with the spin excitations. The infinitesimal positive imaginary part $+i0$ is added to the frequency $\omega$ to be consistent with the Landau causality prescription. Computing the integral of (\ref{SF}) in the complex plane by means of the relation $1/z+i0=P(1/z)+i\pi\delta(z)$, and noticing that the integral over $\varphi$ ($\varphi$ is the angle between $\mathbf{k}$ and $\mathbf{v}$) vanishes for the principal part $P$ as a consequence of symmetric integration limits, we obtain the explicit expression
\begin{equation}
\mathbf{F}_m=\frac{U_m n_0m^2}{12\pi \hbar^3v^2}\left(v+\sqrt{v^2-c_m^2}\right)^3 \mathbf{u}_v.
\label{Force}
\end{equation}
We observe that the force is zero for $v<c_m$ and monotonically increases with $v$ if $v>c_m$ [see Fig.\ref{fig2}(b)]. For overcritical flows, $v\gg c_m$, the force reduces to the classical result $F_m\propto v$. Notice that Eq. (\ref{Force}) is formally similar, but physically different, to the drag force $\mathbf{F}_d=U_d^2 n_0 m^2(v^2-c_s^2)/(\hbar^3 v) \mathbf{u}_v$ acting on the superfluid density in the presence of a localized barrier $V_d=U_d\delta(\mathbf{r})$ \cite{Impurity}, where the role of critical velocity $c_m$ is played by $c_s$.

\emph{Conclusions.---}
We have shown that magnetized spinor-1/2 condensates can independently exhibit density and spin superfluidity as a consequence of the separation between density and spin excitation channels. The different regimes depend on the type of the spin-spin interaction parameter, the applied magnetic field, and the speed of the flow. A feature identified as absolute superfluidity requires the complete and simultaneous suppression of density and spin excitations, a feature that can be observed in a quite large range of parameters. Remarkably, the spin-density separation allow the onset of spin superfluidity even in the supersonic regime. The superfluid properties of a magnetized spinor BEC are summarized in a dynamical phase diagram. A good candidate for the experimental observation of our predictions would be a circularly polarized exciton-polariton condensate either in ferromagnetic regime \cite{Vladimirova} or under an applied magnetic field to compensate the antiferromagnetism \cite{SpinMeissner,PolaritonMagnetic}. Exciton-polaritons constitute, to the best of our knowledge, the most flexible system for experiments with flowing spinor condensates \cite{AmoNaturePhysics}. The observation of superfluid spin currents would represent an important milestone in the fields of spintronics and spin-optronics.

\emph{Acknowledgments.---}
We acknowledge the financial support from the ANR QUANDYDE project and FP7 ITN Spin-Optronics (237252).


\begin{thebibliography}{99}
\bibitem{Note} Weak and higher frequency spin waves are also developping upstream corresponding to the second farther state crossing the energy of the condensate that has a negative group velocity.
\end{thebibliography}
\end{document}